# Hand in Hand Evolution of $\beta$-relaxation and Boson Peak in Metallic Glasses

B. Huang, Z. G. Zhu, T. P. Ge, H. Y. Bai and W. H. Wang*

*Institute of Physics, Chinese Academy of Sciences, Beijing 100190, P. R. China*

**Abstract**

Boson peak and $\beta$-relaxation are two intrinsic and markedly different dynamic modes of glasses, and their structural origins are long-standing issues. Through tuning atomic packing of a model metallic glass with microalloying of different types of elements, we find that low-temperature boson heat capacity peak evolves hand in hand with high-temperature $\beta$-relaxation. A linear correlation between the boson peak temperature and the activation energy of $\beta$-relaxation is disclosed. The coupling of the boson peak and the $\beta$-relaxation indicates their common structural origin of the loosely packed regions in metallic glasses.

## I. INTRODUCTION

All types of glasses display excess specific heat below about 40 kelvin caused by excess low-energy excitations which markedly differ from their crystalline counterparts [1]. The quasilinear temperature ($T$) dependence of specific heat below about 1 K is ascribed to two-level tunneling systems (TLS), and the broad maximum in $T^3$ reduced specific heat at higher $T$ is dominated by the boson peak (BP) vibrational modes over the Debye contribution [1]. As a prominent feature of glasses, The BP has been explained by various theoretical approaches such as soft potential model (SPM) [2,3], fluctuation of force constants on a cubic lattice [4] and vibrational instability of weakly interacting modes [5,6]. However, the origin of BP remains a matter of fierce controversies [7,8]. Recent experiments and theories show that BP may associate with the "defective" loosely packed structure in glasses [9-15]. For example, the localized modes of guest atoms in binary borate glasses overlap with BPs [10]; the progressive decrease of poorly packed boroxol rings caused by pressure leads to reduction of BP for vitreous $B_2O_3$ [11]; with a broadband version of picosecond photo-acoustics technique, BP is related to marginally stable regions in glasses [12]; severely deformed metallic glasses (MGs) show intensified BPs related to the increase of highly localized nano-scaled shear transformed zones (STZs) [13]; theoretical work identifies BP to be hybridized modes between transverse acoustic modes and local librational vibrations in loosely packed nano-regions [14,15].

In addition to primary $\alpha$-relaxation, the secondary $\beta$-relaxation, which emerges near glass transition temperature ($T_g$) [16], is another universal feature of glasses and generally is attributed to reorientational or translational motions of a small number of molecules confined in loosely packed regions [17]. The activation energy of $\beta$-relaxation ($E_\beta$) is nearly equivalent to the energy barrier of STZ indicating that mechanically activated $\beta$-relaxation results in the formation of STZ [18]. MGs have relatively simple atomic structures which can be manifested as dense packing of hard spheres. The $\beta$-relaxation of conventional MGs exists mostly as excess wings or shoulders in dynamical mechanical spectroscopies (DMS) which are difficult to be distinguished from the dominant α-relaxation. Recently, a LaAlNi bulk MG was developed showing a pronounced β-relaxation peak well separated from the α-relaxation in its DMS [19]. The BP of the nonmagnetic MG can be obtained from the low-$T$ heat capacity.

It is intriguing to know if and how the two dynamic modes of BP and β-relaxation, having markedly difference in activation energies and features, connect with each other, and their structural origins. In this letter, we select the $La_{70}Al_{15}Ni_{15}$ MG as a model system to study the relationship of BP and β-relaxation through varying the two dynamic modes by tuning its atomic packing using substitution of solvent La with different types of elements. We show the connection between BP and β-relaxation exists which is caused by their common structural origin of the loosely packed regions.

## II. EXPERIMENTAL AND COMPUTATIONAL DETAILS

Cylindrical amorphous rods of $La_{70-x}Al_{15}Ni_{15}Cu_x$ (x=0, 2, 5, 8, 10), $La_{70-x}Al_{15}Ni_{15}Pd_x$ (x=2, 5), $La_{70-x}Al_{15}Ni_{15}Ti_x$ (x=2, 5), $La_{68}Al_{15}Ni_{15}M_2$ (M=Si, Sn) and $La_{60}Al_{15+x}Ni_{15}Cu_{10-x}$ (x=5, 8, 10) MGs (2 mm in diameter) were fabricated by copper mold casting. We substituted La with Cu, Pd, Ti, Si and Sn as these atoms have different atomic radii. We changed the concentration of Al as the peak temperature of β-relaxation is independent of Al in La-based MGs [20]. The $La_{70-x}Al_{15}Ni_{15}M_x$ and $La_{60}Al_{15+x}Ni_{15}Cu_{10-x}$ MGs are labeled as Mx and AlxCu10-x in the following tables and figures, respectively. DMS were measured on a TA Q800 dynamical mechanical analyzer. Single-cantilever bending method was used with a heating rate of 3 K/min, strain amplitude of 5 $\mu$m and discrete testing frequency of 1, 2, 4, 8, and 16 Hz.

The heat capacity $C_p$ ($C_p \approx C_V$, error<2%) was measured in a physical property measurement system (PPMS 6000) of Quantum Design from 2.0 K to 101 K. The specific heat $C_p$ below 6 K of all the MGs is fitted with equation $C_p=A_{el}T+A_{TLS}T+A_DT^3+A_{sm}T^5=\gamma T+A_DT^3+A_{sm}T^5$ according to SPM, where $A_{el}T$, $A_{TLS}T$, $A_DT^3$ and $A_{sm}T^5$ represents the contributions of electrons, TLS, phonons and excess vibrational modes, respectively [21,22]. Scaled heat capacity $(C_p-\gamma T)/T^3$ in a semilogarithmic scale is plotted in the left part of Fig. 1(a)-(d). Broad peak around 7 K appears for all the MGs. With a harmonic approximation of the free energy, the peak regions are fitted with equation [23]

$$C_p = \gamma T + n_D C_D + 3R \int_0^\infty (\frac{\hbar\omega}{2k_B T})^2 \sinh^{-2}(\frac{\hbar\omega}{2k_B T})[n_{ex} \cdot \chi_{ex}(E_{ex})]dE_{ex}, \quad (1)$$

where $n_D$ and $n_{ex}$ represent the fractions of Debye and excess modes ($n_D+n_{ex}=1$), respectively. The Debye term $n_D C_D$ is contributed by elastic matrix, and $n_D$ equals to 0.88 obtained by dynamic mechanical experiments [19]. $C_D = 9R(\frac{T}{\theta_D})^3 \int_0^{\theta_D/T} \frac{\xi^4 e^\xi}{(e^\xi-1)^2} d\xi$, and the Debye temperature $\theta_D$ is determined by acoustic velocities and density [23]. The obtained $n_D C_D$ is closed to $A_D T^3$ from SPM. To approximately fit the BP, we suppose the Gaussian distribution of the energies of excess modes, i. e., $\chi(E_{ex}) = \frac{1}{\sigma_{ex}\sqrt{2\pi}} e^{-\frac{(E_{ex}-\bar{E}_{ex})^2}{2\sigma_{ex}^2}}$. The fitting $\gamma$, $n_D$, $\theta_D$, $n_{ex}$, $\bar{E}_{ex}$ and $\sigma_{ex}$ are listed in Supplementary Table I, and the fitting curves are shown in Fig. 1(a)-(d) (the solid lines). The Debye contributions are marked by dashed lines and obvious BPs above Debye levels appear around 7 K.

### III. RESULTS AND DISCUSSIONS
#### A. Evolution of boson peak and *β*-relaxation

As shown in the left part of Fig. 1(a), the intensity of the BP of $La_{70}Al_{15}Ni_{15}$ MG monotonically decreases, and the position shifts to higher $T$ with the addition of Cu. The right part of Fig. 1(a) shows $T$ dependent loss modulus $E''$ at 1 Hz scaled by loss modulus $E''_p$ at $T_g$ for the MGs, and separated *β*- and *α*-relaxation peaks appear in the DMS. With the addition of Cu, the *β*-relaxation peak gradually moves to higher $T$. The change tendency coincides with the behavior of the BP. As the atomic radius of Cu ($r_{Cu}$ =1.28 Å) is much smaller than that of La ($r_{La}$ =1.87 Å), the substitution of La with Cu makes the structural cavities of La-formed matrix smaller, thus compacts the loosely packed regions and influences the BP and $\beta$-relaxation. With the substitution of La with Pd [$r_{La}$ > $r_{Pd}$ =1.37 Å> $r_{Cu}$], the BP of La$_{70}$Al$_{15}$Ni$_{15}$ also moves to higher $T$ with the intensity decreasing, and the $\beta$-relaxation peak moves to higher $T$ too. Compared with the same amount of substitution of Cu (*e.g.* 2% or 5%), the substitution effect on the BP and $\beta$-relaxation is weaker due to the smaller reduction of the cavities in La-formed matrix. Unlike Cu addition, the $\alpha$-relaxation peak position significantly varies with the substitution of Pd, which would be caused by strong chemical interaction of Pd with La-formed matrix fully activated at the $\alpha$-relaxation temperature [$\Delta H^{mix}$ of Pd and La is -82 kJ/mol much larger than that of Cu and La (-26 kJ/mol) [24]]. The substitution with demixing Ti ($r_{Ti}$ =1.47 Å) showing a positive $\Delta H^{mix}$ of 20 kJ/mol with La [24] leads to non-monotonic behavior of the BP and $\beta$-relaxation, and the sluggish change of the $\beta$-relaxation as shown in Fig. 1(c). The recovery of BP and $\beta$-relaxation is caused by the looser packing as more excess volume exists with more Ti addition [25]. The covalent bonding substitution of La by Si results in the decrease of the BP, and the shift of the BP and the $\beta$-relaxation peak to higher $T$ [Fig. 1 (d)]. The obvious variation of the BP and $\beta$-relaxation for 2% Si addition is consistent with the small atomic radius ($r_{Si}$ =1.11 Å) of Si.

The $\alpha$-relaxation is expected to be altered only for the addition of element strongly interacting with La (having large negative $\Delta H^{mix}$ with La or tending to form strong bonds). Nevertheless, the $\beta$-relaxation always strongly correlates with the BP regardless of the alloyed elements having different $\Delta H^{mix}$ values with La, being nonmetallic or covalently bonding with La and Ni, and they evolve hand in hand implying their common structural origin. The characteristic of BP induced by the substitution of La with small atoms is almost identical to the dependence of the BP-like hump in specific heat on the relative size of the guest atom to that of the cage for clathrate compounds [26], where the BP-like hump arises from the rattling of small atoms encapsulated in cages forming a lattice with translational invariance [26]. This, together with the independence of BP and $\beta$-relaxation positions on Al content in Fig. 1(d), implies that the BP of LaAlNi MGs could also relate to vibrations of small loose atoms (Ni) encaged in elastic matrix (mainly formed by La). The well fittings with the Gaussian distribution of energies rather than a single energy of excess modes suggests a collective broadband damped rattling due to a set of loose atoms with hierarchical bonds. The loose packed regions in MGs induces the hand in hand evolution of the BP and the $\beta$-relaxation.

**B. Relation of boson peak temperature ($T_B$) and $E_\beta$**

To quantify the correlation of the BP and $\beta$-relaxation, $E_\beta$ is estimated through fitting the $f$ dependent $\beta$-relaxation peak temperature ($T_\beta$) at different frequencies with Arrhenius equation $f = f_\infty \exp(-E_\beta/RT_\beta)$, where $f_\infty$ is the prefactor. The obtained $E_\beta$, position $T_B$ and height [$H_B$, the maximum of $(C_p-\gamma T-C_D)/T^3$] for the BP of these MGs are listed in Table II. Figure 2 plots $E_\beta$ vs. $T_B$, and their linear relations with the additions. The smaller slope for the Ti addition corresponds to the minor change of the $\beta$-relaxation position in Fig. 1(c). According to Eq. (1), $T_B$ is determined by $\bar{E}_{ex}$ and $\sigma_{ex}$. Fig. 2 also shows the dependence of $E_\beta$ on $\bar{E}_{ex}/\sigma_{ex}$, and a similar linear relation is found. $\sigma_{ex}/\bar{E}_{ex}$ (as listed in Table II) reflects fluctuation of the energies of excess modes, influenced by the random location or the distribution of bonding strength of loose atoms, as well as the hybridization of their vibrations with acoustic phonons [26]. Therefore, the dependence of $E_\beta$ on $T_B$ implies the activation of $\beta$-relaxation also associates with the bonding strength of small atoms in loosely packed regions. Denser packing of the soft regions with larger bonding strength and a smaller fluctuation of vibrational energies increases both $T_B$ and $E_\beta$.

**C. Scaling of boson peak with the Debye unit and peak value**

The $(C_p-\gamma T)/T^3$ scaled with the Debye contribution and the peak value $[(C_p-\gamma T)/T^3]_{max}$ are shown in Fig. 3(a) and its inset is for Cu and Ti alloyed MGs. For Cu addition, $(C_p-\gamma T)/C_D$ deviates from the same master curve at the small $T/\theta_D$ regions disclosing the change of loosely packed regions cannot be described by the transformation of elastic continuum [27,28]. The cases are similar to all other substitutions as shown in Fig. 3(b) and (c). The smaller $(C_p-\gamma T)/C_D$ at small $T/\theta_D$ for more Cu addition suggests the quicker densification of the soft regions. For the Ti addition, the larger peak positions of $C_D$ scaled BPs and the narrower $[(C_p-\gamma T)/T^3]_{max}$ scaled BPs indicate the different change of the soft regions compared to that of La-formed matrix resulting in the small slope in Fig. 2.

**D. The anomalous dependence of $E_\beta$ on boson peak height ($H_B$)**

Figure 4(a) shows the dependence of $E_\beta$ on the height of BP, $H_B$, for La-based MGs. The $E_\beta$ seems to be less sensitive to $H_B$ with the alloying of Ti and Al. The $E_\beta$ even does not change when Cu in $La_{60}Al_{15}Ni_{15}Cu_{10}$ MG is substituted by Although $H_B$ decreases obviously. The vibrational densities of states $g(E)/E^2$ and the corresponding Debye levels calculated with the fitting data for $La_{70-x}Al_{15}Ni_{15}Cu_x$ (x=0, 2, 5, 8, 10) and $La_{60}Al_{15+x}Ni_{15}Cu_{10-x}$ (x=2, 5, 8, 10) MGs are shown in Fig. 4(b). The BPs in the $g(E)/E^2$ plots between 3.6 and 4.0 meV resembling those obtained with neutron scattering [29] change similarly to that in Fig. 1(a) and (d). The fraction $N$ of the low-energy excess modes with energies smaller than the average energy $\bar{E}$ [=4.741 meV, as marked in Fig. 4(b)] of the excess modes for $La_{70}Al_{15}Ni_{15}$ MG is calculated and listed in Table II for all MGs. The dependence of $E_\beta$ on $N$ for the La-based MGs is similar to the dependence of $E_\beta$ on $H_B$ as shown in Fig. 4(a) (the open symbols). The dependence of $E_\beta$ on $H_B$ and $N$ discloses the influence of the amount of the softest regions on the activation of $\beta$-relaxation, where the low-energy excess modes primarily originate as revealed by simulations [30,31].

Contrary to the dependence of the BP on density $\rho$ in the Cu alloying process, the BP (or $H_B$, $N$) decreases with $\rho$ decreasing when Cu in $La_{60}Al_{15}Ni_{15}Cu_{10}$ MG is substituted by Al [the inset of Fig. 4(b)], suggesting the change of BP and $\beta$-relaxation cannot be simply explained with the density change of the whole sample [27].

**E. Diagram of dynamics in MGs**

A diagram on the relation between BP and $\beta$-relaxation of MGs within experimental time window is constructed based on the loosely packed regions as illustrated in Fig. 5. The BP, $\beta$-relaxation and $\alpha$-relaxation peaks appear at different temperatures [8]. The BP is mainly associated with the quasilocalized vibrations of small atoms (cyan spheres) in the soft regions surrounded by the elastic matrix (loose Ni atoms in La-formed matrix in LaAlNi MG) as depicted by the insert of schematic cages in the diagram. With $T$ approaching $T_g$ the atoms in the soft regions will move cooperatively and reversibly with a large scale, and the cages are deformed, resulting in diffusion or jump of atoms from one cage to another and the occurrence of the $\beta$ event as depicted in the diagram [16,18]. In Cu, Pd, Si and Sn substituting process, due to the reduction of the cavities, the soft regions are more densely packed showing with the decrease of $H_B$ and the increase of $T_B$. Meanwhile, the $E_\beta$ becomes larger, as only under larger thermal fluctuations with sufficient movement of atoms in the soft regions can the activation of $\beta$-relaxation composing of series of $\beta$ events occurs. For Ti substitution, due to its demixing effect with La and the increase of excess volume in the elastic matrix, the $\beta$-relaxation evolves sluggishly compared with BP and both BP and $\beta$-relaxation appear at lower energies with sufficient addition of Ti, although $r_{Ti}$ is smaller than $r_{La}$. With the substitution of Cu with Al, the cage-like structure will not be changed, and the $T_B$ and $E_\beta$ keep constant. Induced by the strong interaction of Al with small atoms Ni ($\Delta H^{mix}$ of Al and Ni is -22 kJ/mol [24]) and the reduction of the cage-like structure, the BP (or $H_B$, $N$) of the MG decreases.

## IV．CONCLUSION

We reveal the connection between BP and $\beta$-relaxation, and demonstrate both the two dynamic modes associate with the nano-scale loosely packed regions or structural heterogeneity in MGs, i. e. they have common structural origin. Our results might shed light on the mechanisms of BP and $\beta$-relaxation, and indicate the strong relation between fast BP dynamics and slow structural relaxation in MGs.

**ACKNOWLEDGEMENTS**

We are grateful for the experimental assistance or discussions of H. L. Peng, D. Q. Zhao, D. W. Ding and S. K. Su. The financial support of the NSF of China (No. 51271195) is appreciated.

———————————————————————

**Table I** The fitting $\gamma$, $n_D$, $\theta_D$, $n_{ex}$, $\bar{E}_{ex}$, $\sigma_{ex}$ of the $C_p$ with Eq. (1) for $La_{70-x}Al_{15}Ni_{15}Cu_x$ (x=0, 2, 5, 8, 10), $La_{70-x}Al_{15}Ni_{15}Pd_x$ (x=2, 5), $La_{70-x}Al_{15}Ni_{15}Ti_x$ (x=2, 5), $La_{68}Al_{15}Ni_{15}Si_2$, $La_{60}Al_{15+x}Ni_{15}Cu_{10-x}$ (x=5, 8, 10) and $La_{68}Al_{15}Ni_{15}Sn_2$ MGs.

| BMGs | $\gamma$ (mJ/mol K$^2$) | $n_D$ | $\theta_D$ (K) | $n_{ex}$ | $\bar{E}_{ex}$ (meV) | $\sigma_{ex}$ (meV) |
|---|---|---|---|---|---|---|
| Cu0 | 4.347 | 0.881 | 162.0 | 0.119 | 4.741 | 1.397 |
| Cu2 | 4.310 | 0.881 | 164.0 | 0.119 | 4.802 | 1.405 |
| Cu5 | 4.467 | 0.881 | 168.0 | 0.119 | 4.905 | 1.422 |
| Cu8 | 4.352 | 0.881 | 171.0 | 0.119 | 5.000 | 1.440 |
| Cu10 | 4.638 | 0.881 | 172.0 | 0.119 | 5.043 | 1.448 |
| Pd2 | 4.275 | 0.881 | 162.3 | 0.119 | 4.741 | 1.397 |
| Pd5 | 4.269 | 0.881 | 165.7 | 0.119 | 4.819 | 1.406 |
| Ti2 | 7.183 | 0.834 | 172.1 | 0.166 | 5.362 | 1.465 |
| Ti5 | 6.015 | 0.865 | 166.3 | 0.135 | 5.017 | 1.414 |
| Si2 | 4.404 | 0.881 | 164.2 | 0.119 | 4.819 | 1.414 |
| Al5Cu5 | 4.447 | 0.888 | 176.8 | 0.112 | 5.043 | 1.465 |
| Al8Cu2 | 4.309 | 0.892 | 177.7 | 0.108 | 5.043 | 1.448 |
| Al10Cu0 | 4.503 | 0.893 | 178.6 | 0.107 | 5.043 | 1.457 |
| Sn2 | 4.734 | 0.881 | 163.7 | 0.119 | 4.8016 | 1.405 |

**Table II** The boson heat capacity peak temperature $T_B$, the height $H_B$, the $E_\beta$ of $\beta$-relaxation, the energy fluctuation $\sigma_{ex}/\overline{E}_{ex}$ and the fraction $N$ of excess soft modes with energies smaller than $\overline{E}$ of 4.741 meV for $La_{70-x}Al_{15}Ni_{15}Cu_x$ (x=0, 2, 5, 8, 10), $La_{70-x}Al_{15}Ni_{15}Pd_x$ (x=2, 5), $La_{70-x}Al_{15}Ni_{15}Ti_x$ (x=2, 5), $La_{68}Al_{15}Ni_{15}Si_2$, $La_{60}Al_{15+x}Ni_{15}Cu_{10-x}$ (x=5, 8, 10) and $La_{68}Al_{15}Ni_{15}Sn_2$ MGs.

| BMGs | $T_B$ (K) | $H_B$ (mJ/mol K$^4$) | $E_\beta$ (kJ/mol) | $\sigma_{ex}/\overline{E}_{ex}$ | $N$ (%) |
|---|---|---|---|---|---|
| Cu0 | 7.1±0.1 | 0.598±0.005 | 88±2 | 0.2946 | 5.946 |
| Cu2 | 7.3±0.1 | 0.562±0.005 | 98±3 | 0.2926 | 5.742 |
| Cu5 | 7.6±0.1 | 0.521±0.005 | 107±4 | 0.2900 | 5.401 |
| Cu8 | 7.9±0.1 | 0.488±0.005 | 115±6 | 0.2879 | 5.099 |
| Cu10 | 8.0±0.1 | 0.474±0.005 | 117±8 | 0.28718 | 4.9651 |
| Pd2 | 7.1±0.1 | 0.588±0.005 | 90±2 | 0.2946 | 5.946 |
| Pd5 | 7.4±0.1 | 0.553±0.005 | 104±3 | 0.2916 | 5.684 |
| Ti2 | 9.0±0.1 | 0.522±0.005 | 94±3 | 0.2733 | 5.575 |
| Ti5 | 8.1±0.1 | 0.534±0.005 | 91±2 | 0.2818 | 5.703 |
| Si2 | 7.3±0.1 | 0.557±0.005 | 98±3 | 0.2934 | 5.686 |
| Al5Cu5 | 7.8±0.2 | 0.452±0.005 | 115±7 | 0.2906 | 4.683 |
| Al8Cu2 | 8.0±0.1 | 0.430±0.005 | 117±6 | 0.2872 | 4.506 |
| Al10Cu0 | 7.9±0.1 | 0.429±0.005 | 114±5 | 0.2889 | 4.469 |
| Sn2 | 7.3±0.2 | 0.562±0.005 | 102±3 | 0.2926 | 5.742 |

**Figure captions**

**Figure. 1** The left part of each figure: $T$ dependent $(C_p-\gamma T)/T^3$ in a semilogarithmic scale for (a) $La_{70-x}Al_{15}Ni_{15}Cu_x$ (x=0, 2, 5, 8, 10), (b) $La_{70-x}Al_{15}Ni_{15}Pd_x$ (x=0, 2, 5), (c) $La_{70-x}Al_{15}Ni_{15}Ti_x$ (x=0, 2, 5), (d) $La_{70-x}Al_{15}Ni_{15}Si_x$ (x=0, 2) and $La_{60}Al_{15+x}Ni_{15}Cu_{10-x}$ (x=0, 5, 8, 10) MGs. The solid lines are the fitting lines. The dashed lines are the Debye contributions. The right part of each figure shows corresponding $T$-dependent loss modulus $E''$ scaled by the $E''_p$ at $T_g$ with 1 Hz.

**Figure. 2** The activation energy $E_\beta$ of $\beta$-relaxation versus the peak temperature of BP $T_B$ and $\bar{E}_{ex}/\sigma_{ex}$ (the open symbols) for the La-based MGs alloyed with Cu, Pd, Ti, Si, Al and Sn. The solid and dashed lines are the guides to the eyes.

**Figure. 3** The $(C_p-\gamma T)/C_D$ versus $T/\theta_D$ with the inset showing BPs scaled by the peak value for (a) the Cu and Ti alloyed MGs; (b) the Pd, Si and Sn alloyed MGs; (c) the Al alloyed MGs.

**Figure. 4 (a)** The $E_\beta$ versus the height $H_B$ of the BP [the maximum of $(C_p-\gamma T-C_D)/T^3$] and $N$ (the open symbols) for the La-based MGs alloyed with Cu, Pd, Ti, Si, Al and Sn. The solid and dashed lines are the guides to the eyes. (b) The deceasing BPs in vibrational density $g(E)/E^2$ for $La_{70-x}Al_{15}Ni_{15}Cu_x$ (x=0, 2, 5, 8, 10) and $La_{60}Al_{15+x}Ni_{15}Cu_{10-x}$ (x=5, 8, 10) MGs with the inset (the Cu and Al labels represent the Cu and Al alloying MGs, respectively) showing the density $\rho$. The dash lines mark the Debye levels. The dashed lines with arrows in the inset are the guide to the eyes.

**Figure 5.** $T$ dependent glassy dynamics in the alloying process. The diagrams above the BPs and $\beta$-relaxation peaks depict the quasilocalized vibrations and the $\beta$ event in the soft regions.

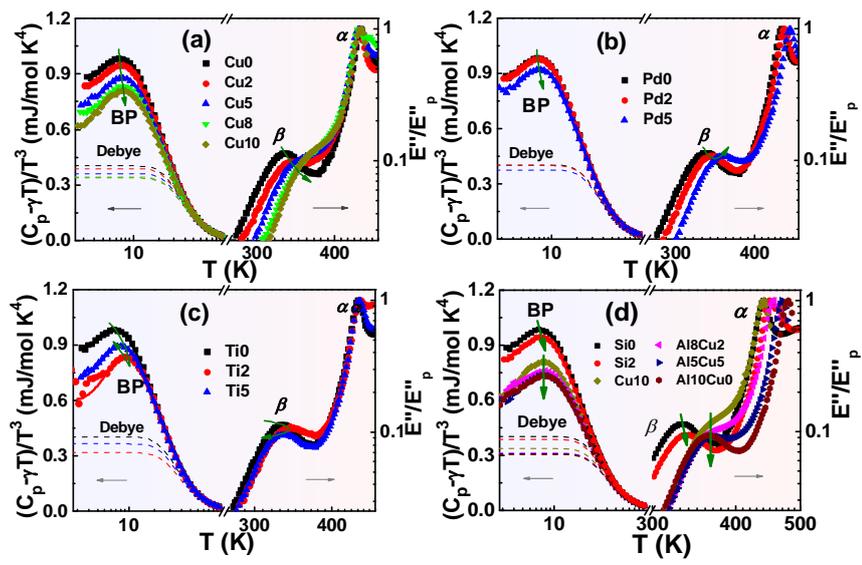

Figure 1, Huang, *et al.*

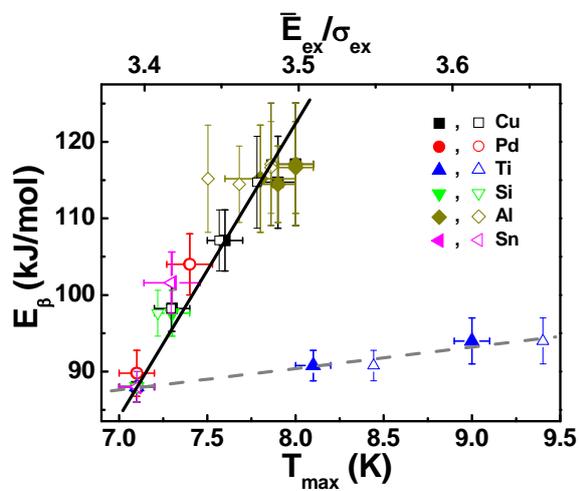

Figure 2, Huang, *et al.*

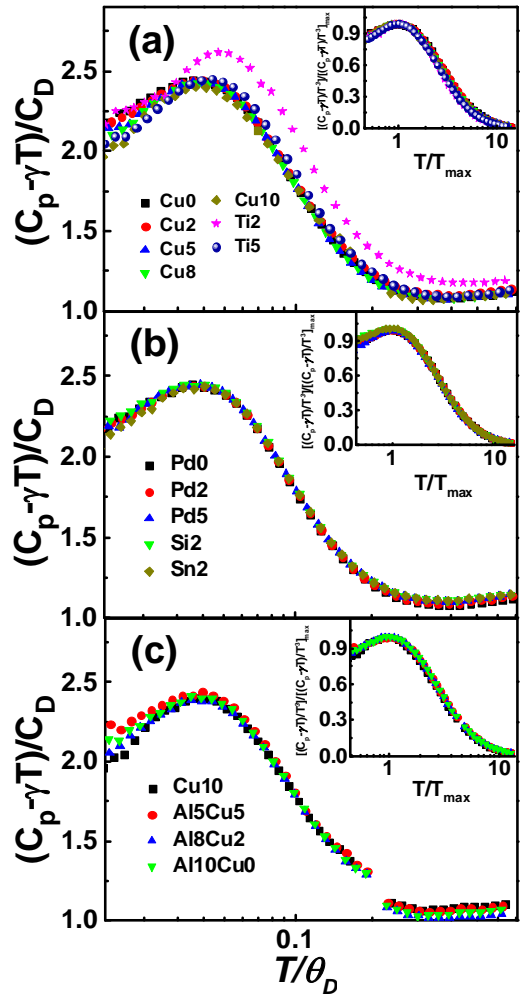

Figure 3, Huang, *et al.*

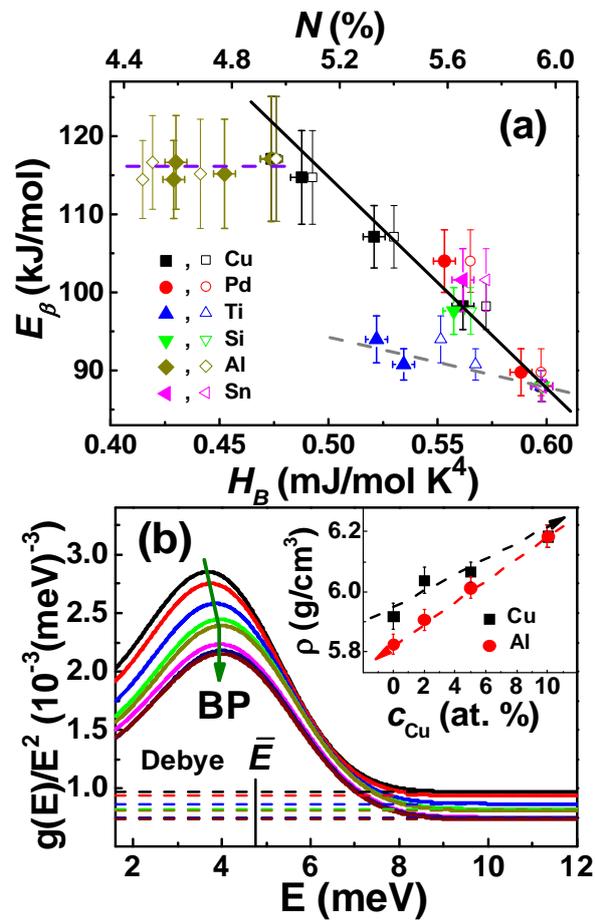

Figure 4, Huang, *et al.*

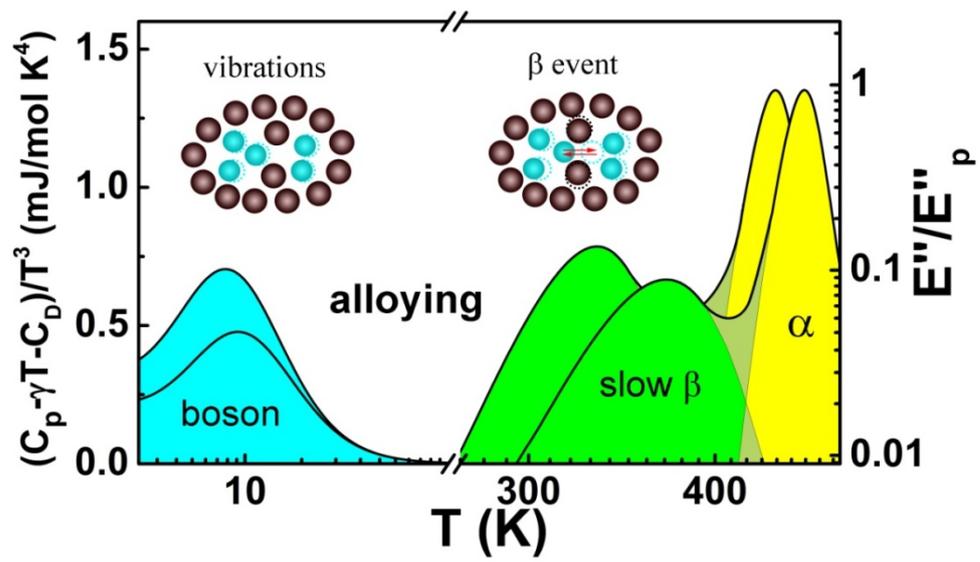

Figure 5, Huang, *et al.*